                                 \documentstyle[12pt,aaspp4]{article}
                                \begin{document}

                                \title{
Constraints on Type Ia Supernova Models\\
from X-ray Spectra of Galaxy Clusters
                                }

                                \author{
Renato A. Dupke \& Raymond E. White III
                                }

                                 \affil{
University of Alabama, Tuscaloosa, AL 35487-0324
                                }

                                \begin{abstract}
We present constraints on theoretical models of Type Ia supernovae using
spatially resolved {\sl ASCA} X-ray spectroscopy of three galaxy clusters:
Abell 496, Abell 2199 and Abell 3571.
All three clusters have central iron abundance enhancements;
an ensemble of abundance ratios are used to show that most of the
iron in the central regions of the clusters comes from SN Ia.
These observations are consistent with the suppressed
galactic wind scenario proposed by Dupke \& White (1999).
At the center of each cluster,
simultaneous analysis of spectra from all {\sl ASCA} instruments
shows that the nickel to iron abundance ratio (normalized by the solar
ratio) is Ni/Fe $\approx4$.
We use the nickel to iron ratio as a discriminator between
SN Ia explosion models:
the Ni/Fe ratio of ejecta from the ``Convective Deflagration'' model W7
is consistent with the observations, while those of ``delayed detonation''
models are not consistent at the 90\% confidence level.
                                \end{abstract}

                                \clearpage
                                \section{
Introduction
                                }

Type Ia supernovae are thought to be generated by thermonuclear explosions of
carbon-oxygen white dwarfs undergoing accretion in stellar binary systems
(Hoyle \& Fowler 1960).
However, the nature of the progenitor binary systems (one white dwarf or two?),
the masses of the white dwarfs (Chandrasekhar or sub-Chandrasekhar mass?)
and the explosion mechanism(s) (e.g. convective deflagration,
delayed-detonation, etc.) for SN Ia are still
open questions (for recent reviews see Branch 1998; Branch et al.\ 1995;
Niemeyer \& Woosley 1997; Nomoto, Iwamoto \& Kishimoto 1997;
Ruiz-Lapuente et al.\ 1997).

The explosion of a Chandrasekhar mass C-O white dwarf is thought to be
initiated by carbon ignition at the center, followed by a subsonic
nuclear flame (deflagration wave) propagating outwards.
Prompt detonation models (Arnett 1969; Hansen \& Wheeler 1969),
characterized by supersonic nuclear flames, are not viable since they do not produce
the intermediate mass elements which are observed in SN Ia spectra
(Nomoto, Thielemann, \& Yokoi 1984; Wheeler \& Harkness 1990).
Among deflagration (subsonic nuclear flame) models, the elemental nucleosynthesis
depends on the characteristics of the flame propagation, on shock waves generated by
instabilities at the flame front and on the density at the transition from deflagration
to detonation (Nomoto et al.\ 1997b; Niemeyer \& Hillebrandt 1995; Niemeyer \& Woosley 1997).
In the classical ``W7'' deflagration model
(Nomoto et al.\ 1984; Branch et al.\ 1995; Harkness 1991; Thielemann, Nomoto \& Yokoi 1986),
the propagation speed of the flame front is relatively high ($\sim 15-25\%$ of the
sound speed), but remains subsonic.
In ``delayed detonation'' models, the flame speed is initially much lower,
($\sim 1-3\%$ of the sound speed), but rises to become supersonic
(Livne 1993; Arnett \& Livne 1994a, 1994b; Khokhlov 1995; Niemeyer \& Hillebrandt 1995).
In  ``pulsating detonation'' models, the deflagration fails to unbind the white dwarf,
causing a pulsation, and detonation occurs upon contraction (Arnett \& Livne 1994a, 1994b).

Despite the remaining theoretical uncertainties in the more popular
models for SN Ia, there is
better agreement in their predicted nucleosynthetic yields than in the
yields from various models for SN II (Gibson, Loewenstein \& Mushotzky 1997).
Nonetheless, significant discrepancies still exist between the yield predictions of
competing SN Ia explosion models.
For example, the nickel yield from the W7 model cited above is $\sim3$ times
greater than those from delayed detonation models.
Accurately determining the elemental yields from SN Ia is crucial to determining the
relative contribution of different supernovae types to the metal enrichment of galaxies
and intracluster gas. Furthermore, assessing the relative contribution of SN Ia and II
ejecta in intracluster gas provides crucial information for determining which
enrichment mechanism(s) were most dominant in contaminating intracluster gas.
In this paper we use X-ray spectroscopic observations of intracluster gas
to discriminate between competing theoretical models for SN Ia.

Intracluster gas tends to be metal-rich, with abundances of $\sim0.3-0.4$ solar.
Since heavy elements are produced in supernovae,
the metals in intracluster gas must have come from stars.
However, the metal enrichment mechanism(s) for intracluster gas remains controversial.
The most likely enrichment mechanisms are thought to be protogalactic winds
(Larson \& Dinerstein 1975) and
ram pressure stripping (Gunn \& Gott 1972).
The processes may be distinguished by their chemistry.
Protogalactic winds would be associated with SN II ejecta, while the more secular
process of ram pressure stripping would be associated with significant
supplemental amounts of SN Ia ejecta which accumulated in the interstellar
medium of galaxies.

Early X-ray spectroscopy of intracluster gas indicated a dominance of SN II ejecta
(Canizares et al.\ 1982; Canizares, Markert \& Donahue 1988) and more recent {\sl ASCA}
spectroscopy has been interpreted similarly
(Mushotzky \& Loewenstein 1997;  Mushotzky et al.\ 1996).
Such a dominance of SN II ejecta supports the protogalactic wind scenario for the
metal enrichment of intracluster gas.
Further support comes from the observation that the specific
energy of intracluster gas is greater than that of cluster galaxies (White 1991).
However, theoretical uncertainties in the elemental yields from SN Ia and SN II
allow {\sl ASCA} spectroscopy to also be interpreted as showing that as much as
50\% of the iron in clusters comes from SN Ia
(Ishimaru \& Arimoto 1997; Fukazawa et al.\ 1998; Nagataki \& Sato 1998;
Dupke 1998; Dupke \& White 1999).
In a detailed study of Abell 496, Dupke \& White (1999) used an ensemble of elemental
abundance ratios to show that the fractional contribution of SN Ia ejecta is
$\sim50\%$ (by mass) in the bulk of the cluster, increasing to $\sim70\%$ in the
vicinity of its central cD galaxy.
They argued that ram pressure stripping could not have caused the
central abundance enhancement in Abell 496 and instead proposed that a secondary
SN Ia-driven wind (following a more vigorous SN II-driven protogalactic wind),
was partially suppressed in the vicinity of the cD (due to the cD being
at the bottom of the cluster's gravitational potential
and in the midst of the highest ambient intracluster gas density).
If this suggestion is correct, high fractions of SN Ia ejecta should
also be found in other clusters with central abundance enhancements.

In this paper we show that Abell 496, Abell 2199, and Abell 3571
have central iron abundance enhancements.
We use an ensemble of abundance ratios to determine the relative
proportion of SN Ia and SN II ejecta in their central regions and find
that the mass fraction of SN Ia ejecta ranges from $\sim60-70\%$.
We then further analyze these central regions, where the metal
contamination is dominated by SN Ia ejecta,
to discriminate between competing theoretical models for SN Ia.
In particular, we show that the Ni/Fe ratios observed at the centers
of these clusters are consistent with the ``standard'' W7 model,
but are inconsistent with those of various delayed-detonation models.

                                \section{
Cluster Characteristics
                                }

Abell 496 is a Bautz-Morgan Type I cluster with an optical redshift of $z=0.0328$.
Adopting a Hubble constant of 50 km s$^{-1}$ Mpc$^{-1}$ and $q_0=0.5$, its
luminosity distance is $197\ h_{50}^{-1}$ Mpc and $1'=57\ h_{50}^{-1}$ kpc.
Neither the projected galaxy distribution nor the galaxy velocity distribution
in the cluster shows signs of significant substructure, so the cluster appears to be
dynamically relaxed (Bird 1993; Zabludoff, Huchra \& Geller 1990).
Nulsen et al.\ (1982) found a soft X-ray component in $Einstein$ SSS spectra
of this cluster and estimated a cooling accretion rate of
$\sim$200 $M_\odot$ yr$^{-1}$, which is
consistent with later analyses (Mushotzky  1984; Mushotzky
\& Szymkowiak 1988; Canizares et al.\ 1988;
Thomas, Fabian \& Nulsen 1987; White et al.\ 1994).
In the course of a joint analysis of $Einstein$ SSS and $Ginga$ LAC spectra,
White et al.\ (1994) found a central abundance enhancement in Abell 496, which
was later confirmed by {\sl ASCA} observations (Dupke \& White 1999).
Dupke \& White (1999) also found gradients in elemental abundance $ratios$,
indicative of a spatial gradient in the proportion of SN Ia/II ejecta.

Abell 2199 is a Bautz-Morgan Type I cluster with an optical redshift $z = 0.0309$.
Its luminosity distance is $185\ h_{50}^{-1}$ Mpc and $1'=54\ h_{50}^{-1}$ kpc.
This cluster is among the 10 X-ray brightest galaxy clusters
(Edge \& Stewart 1991; Siddiqui, Stewart \& Johnstone 1998).
The cluster X-ray emission is well centered at its central cD galaxy, NGC 6166.
Neither the projected galaxy distribution nor the galaxy velocity distribution
shows signs of significant substructure (Buote \& Tsai 1996).
Previous X-ray analyses of this cluster with {\sl EXOSAT} (Edge 1989), {\sl Einstein}
(Stewart et al.\ 1984; Thomas et al.\ 1987), {\sl Ginga/Einstein} (White et al.\ 1994),
and {\sl ROSAT} (Allen \& Fabian 1997; Siddiqui et al.\ 1998) indicate the presence of a
cooling flow with an accretion rate of $\dot M \approx 80 - 250$ $M_{\odot}$ yr$^{-1}$.
In the course of a joint analysis of {\sl Einstein} SSS and {\sl Ginga} LAC spectra,
White et al.\ (1994) found weak evidence for a central abundance enhancement in Abell 2199.
{\sl ASCA} observations of the external regions ($>3'$) of Abell 2199 have been
previously analyzed by Mushotzky et al.\ (1996).

Abell 3571 (SC 13444-325) is a bright (Lahav et al.\ 1989), moderately rich (richness
class 2) Bautz-Morgan Type I cluster with an optical redshift $z = 0.0397$.
Its luminosity distance is $238\ h_{50}^{-1}$ Mpc and $1'=69\ h_{50}^{-1}$ kpc.
Quintana \& de Souza (1993) suggested that the galaxy positions and velocities
indicate the presence of several subgroups.
However, {\sl ASCA} observations of this cluster do not show signs of significant
substructures within a radius up to $35'$, which indicates that the bulk
of the cluster mass is virialized (Markevitch et al.\ 1998).
Abell 3571 is possibly a member of the Shapley 8 supercluster, which is itself
centered on Abell 3558.
Analysis of {\sl EXOSAT} data indicated
a cooling flow with $\dot M\approx100$ $M_{\odot}$ yr$^{-1}$ (Edge, Stewart \& Fabian 1992).

                                \section{
Data Reduction \& Analysis
                                }

The {\sl ASCA} satellite has four large-area X-ray telescopes, each coupled to its own
detector: two Gas Imaging Spectrometers (GIS) and two Solid-State Imaging Spectrometers (SIS).
Each GIS has a $50'$ diameter circular field of view and a usable energy range of 0.7--12 keV;
each SIS has a $22'$ square field of view and a usable energy range of 0.4--10 keV.

Both Abell 496 and Abell 2199 were observed for 40 ksec by {\sl ASCA} in September
and July of 1993, respectively.
Abell 3571 was observed for 30 ksec in  August 1994.
For all three clusters we employed standard reduction procedures,
selecting data taken with high and medium bit rates, with cosmic ray
rigidity values $\ge$ 6 GeV/c, with elevation angles from the
bright Earth of $\ge20^{\circ}$ and from the
Earth's limb of $\ge5^{\circ}$ (GIS) or $10^{\circ}$ (SIS), and we
excluded times when the satellite was affected by the South Atlantic Anomaly.
Rise time rejection of particle events was performed on GIS data and
SIS data had hot and flickering pixels removed.
The resulting effective exposure times for each instrument are listed in Table 1.
We estimated backgrounds from blank sky files provided by the {\sl ASCA} Guest
Observer Facility.

We used XSPEC 10 (Arnaud 1996) software to analyze the
{\sl ASCA} spectra for these clusters.
Spectra were fit using the {\tt mekal} and {\tt vmekal}
thermal emission models, which are based on the emissivity calculations of Mewe \& Kaastra
(cf. Mewe, Gronenschild \& van den Oord  1985; Mewe, Lemen \& van den Oord 1986;
Kaastra  1992), with Fe L calculations by Liedahl, Osterheld \& Goldstein (1995).
Abundances are measured relative to the solar photospheric values of
Anders \& Grevesse (1989), in which Fe/H=$4.68\times10^{-5}$ by number.
Galactic photoelectric absorption was incorporated using the {\tt wabs}
model (Morrison \& McCammon  1983).
Spectral channels were grouped to have at least 25 counts/channel.
Energy ranges were restricted to 0.8--10 keV for the GIS and 0.4--10 keV for
the SIS.

Spectra from all four {\sl ASCA} instruments (SIS 0 \& 1 and GIS 2 \& 3)
were fit individually and jointly.
Since the individual spectral fits are consistent with the joint
analyses of all four instruments, we will only describe the joint fits.
Spectra were extracted from regions as small as $2'$ radius at several
different projected spatial distances from the clusters center.
Thermal emission models ({\tt mekal}) with variable temperatures,
overall abundances, normalizations and absorbing column densities ({\tt wabs})
were then jointly fit to all four spectra from each region.
The normalizations for the individual spectra were allowed to vary
independently, to compensate for small calibration and spatial extraction
differences between the four detectors.
Since each cluster has a moderate cooling flow at its center,
we also added a cooling flow component to the {\tt mekal} thermal emission
model for the central region of each cluster, in order to test the
model-dependence of our abundance measurements.
The cooling flow spectral model {\tt cflow} in XSPEC is characterized by
maximum and minimum temperatures, an abundance, a slope which parameterizes
the temperature distribution of emission measures, and a
normalization which is simply the cooling accretion rate.
We adopted the emission measure temperature distribution that corresponds
to isobaric cooling flows (zero slope).
We tied the maximum temperature of the cooling flow to the temperature of the
thermal component, and we fixed the minimum temperature at 0.1 keV.
We applied a single (but variable) global absorption to both spectral
components and associated an additional, intrinsic absorption component with
the cooling flow, placing it at the redshift of the cluster.
The resulting fits were all excellent, having reduced $\chi^2$ of $\chi^2_\nu\approx1$.

                                        \section{
Results
                                         }

                                        \subsection{
Abundance and Temperature Distributions
                                        }

We first used isothermal {\tt mekal} models to determine the overall elemental
abundance (largely driven by iron) in each cluster region.
Radial distributions of these elemental abundances are shown in Figure 1
and listed in Table 2;
the indicated errors are 90\% confidence limits.
Results for two projected spatial regions are shown, a central region
from $0-2'$ and an outer region extending from $3-12'$.
All three clusters show mild, but significant, abundance enhancements at the center.
For Abell 496, the abundance is 0.53$^{+0.04}_{-0.04}$ solar at the center, falling
to 0.36$^{+0.03}_{-0.03}$ solar in the outer $3-12'$ region (Dupke \& White 1999).
In Abell 2199, the abundance declines from a central value of $0.49 \pm 0.04$ solar
to $0.34 \pm 0.03$ solar in the outer parts.
In Abell 3571, the central abundance is $0.37\pm 0.06$ solar, declining
to $0.28\pm 0.03$ solar in the outer regions.

We also used the $F$-test to assess the significance of the abundance gradient
in each cluster by comparing the $\chi^2$ of fits which assumed the abundances
were the same in the two projected spatial regions (inner versus outer) to fits which
allowed the abundances in the two regions to vary independently.
The difference between the $\chi^2$ of these two fits must follow a
$\chi^2$ distribution with one degree of freedom (Bevington 1969).
We find that the central abundance enhancements are significant with $>99.99$\%
confidence.
As can be seen from Table 2, the addition of a cooling flow component
in the central regions does not significantly change the abundance estimates.

Both Abell 496 and Abell 2199 have clear temperature gradients when
isothermal {\tt mekal} models are used:
the temperature in Abell 496 rises from $3.24_{ -0.06}^{+0.07}$ keV within $2'$ to
$4.40_{-0.13}^{+0.13}$ keV beyond $3'$;
in Abell 2199 the temperature rises from a central value of
$3.7 \pm 0.09$ keV to $4.14\pm0.06$ keV beyond $3'$.
These results are consistent with those of Mushotzky et al.\ (1996).
The average temperature of the gas in Abell 3571 is $6.75 \pm 0.16$ keV,
which is consistent with the results of Markevitch et al.\ (1998),
and the gradient within $12'$ is not significant at the 90\% confidence level.
For models including a cooling flow component at the center, the temperature
gradients in Table 2 $appear$ much weaker than described above, since
we quote only the best-fitting initial (upper) temperature of the cooling gas.

                                \subsection{
Individual Elemental Abundances and Abundance Ratios
                                }

We also determined individual elemental abundances in the central region of
each cluster by using the {\tt vmekal} spectral model in XSPEC.
In our spectral model fits, the helium abundance was fixed at the solar value, while
carbon and nitrogen were fixed at 0.3 solar (since {\sl ASCA} is rather insensitive
to carbon and nitrogen and the derived abundances of other elements
are not affected by the particular choice for carbon and nitrogen abundances).
Our observed abundances are shown in Table 3.

It can be seen from Table 3 that the best determined abundances in all three
clusters are those of iron, silicon and nickel.
Argon, calcium and magnesium abundances are poorly measured and are not listed.
Oxygen, neon and sulfur abundances are not well constrained in Abell 3571.
Nickel abundances are about twice solar in all three clusters.

To estimate the SN Ia/II mass fraction, we will compare various observed abundance
ratios to the theoretical predictions of specific models for SN Ia and II.
For SN Ia we initially adopt the updated W7 model of Nomoto et al.\ (1997b),
while for SN II we use the calculations of Nomoto et al.\ (1997a, 1997b), who adopt
a Salpeter initial mass function over a SN II progenitor mass range of 10-50 $M_\odot$.
The abundance ratios we use involve oxygen, silicon, iron and nickel.
Various observed ratios are listed in Table 4 and compared to the theoretical
ratios from the models we adopt.

We do not include sulfur and neon in these estimates because their theoretical
yields from SN II are particularly problematic.
In their analysis of {\sl ASCA} spectra of four clusters, Mushotzky et al.\ (1996)
observed a persistent underabundance of sulfur and an overabundance of
neon compared to expectations from theoretical models of supernova yields.
Dupke \& White (1999) and Dupke \& Arnaud (1999), using different theoretical
models for SN II yields, found that the theoretical sulfur yield would have to be
reduced by a factor of $\sim$2-4 to be consistent with results for most other
elemental ratios in their {\sl ASCA} analyses of Abell 496 and Perseus, respectively.
Dupke \& Arnaud (1999) also found that theoretical neon yields
would have to be reduced by a factor of $\sim2.7$ to be consistent with
the results for most other elemental ratios.
Therefore, to estimate the SN Ia iron mass fraction, we use only abundance ratios
involving the better constrained (observationally and theoretically) abundances of
iron, silicon, nickel and oxygen.

Various abundance ratios in the central regions of Abell 496, Abell 2199,
and Abell 3571 are shown in Table 4, along with the theoretical
expectations for SN Ia and SN II ejecta; the errors associated with
the observed abundance ratios are the propagated 1$\sigma$ errors.
In Table 5 we list the best-constrained abundance ratios for each cluster
along with the associated estimates for the SN Ia iron mass fractions;
the  individual SN Ia iron mass fraction estimates are consistent with one
another, within their errors.
The ensemble average of these individual estimates of the SN Ia iron
mass fraction (weighted by their errors) is also indicated at the bottom of
Table 5 for each cluster.
(In Abell 3571, the oxygen abundance is not well constrained, so
we have not included it in the calculation of the ensemble average.)
Evidently, the central regions of all three clusters are substantially enriched by
SN Ia ejecta, with SN Ia iron mass fractions of 73$\pm 5\%$, 74$\pm 7\%$ and
58$\pm 17\%$ in Abell 496, Abell 2199 and Abell 3571, respectively.

                           \section{
Comparison to Alternative Models for SN Ia
                                }

In the previous section we showed that Abell 496, Abell 2199 and Abell 3571 have
significant central abundance enhancements which are comprised mostly of SN Ia ejecta.
A variety of abundance ratios provide mutually consistent estimates of
the central SN Ia iron mass fraction, given our adopted models for
theoretical yields from SN Ia and II.
This mutual consistency encourages us to test alternative theoretical
models for SN Ia (alternative SN II models will be explored in White 1999).
In particular, we will compare the W7 ``deflagration'' model adopted above
to ``delayed detonation'' models which have been explored in response to
concerns about the physicality of the explosion in the W7 model (Khokhlov 1991).
Delayed detonation models are characterized by (initially) slower
flame speeds (a few percent of the sound speed) than the W7 model.

We consider the delayed detonation models of Nomoto et al.\ (1997b),
who calculate yields for a sequence of models distinguished by a variety of
densities ahead of the deflagration front, which modulates the onset of detonation.
We find that the elemental yields of their models WDD1, WDD2 and WDD3 do
$not$ provide the same consistency as found with model W7 in estimating the
SN Ia/II fraction.
Table 6 compares the estimates of SN Ia iron mass fractions derived from
the three delayed detonation models described above.
The adopted SN II yields are again those of Nomoto et al.\ (1997a, 1997b).
Comparison of Tables 5 and 6 shows that the W7 model provides more satisfactory
results in three different ways: 1) more abundance ratios fall within the theoretical
bounds of SN Ia and SN II, as they should, when W7 is adopted
(observed ratios which are outside the theoretical bounds for SN Ia and II ejecta,
therefore preventing mass fraction estimates, are indicated
by dashes); 2) the dispersion in the average of the (remaining) individual
SN Ia iron mass fraction estimates tends to be smaller when the W7
model is adopted; 3) the mutual consistency of the SN Ia iron mass fraction
estimates derived from individual abundance ratios for a given cluster
is best when the W7 model is adopted.

Concentrating on the nickel yield in particular, we find that our cluster
observations further discriminate against the delayed detonation models
cited above.
The observationally best determined abundance ratio involving nickel is
the nickel to iron ratio.
Recall from Table 4 that the individual Ni/Fe ratios are consistent
with the other indicated ratios in estimating the fractional SN Ia contamination.
Figure 2 compares the Ni/Fe ratio observed in the three clusters analyzed in this
work to the predicted Ni/Fe ratio from the W7 deflagration model, as well as from
the three delayed detonation models described above;
the indicated observational errors are the propagated 90\% confidence limits.
Evidently, the observed Ni/Fe ratios are $inconsistent$ with any of these delayed
detonation models at better than 90\% confidence.
The observed Ni/Fe ratio is fully consistent with the W7 model, however.
Thus, the elemental yields of the W7 model are more consistent with
X-ray spectroscopic observations of intracluster gas than those of delayed
detonation models.

                           \section{
Summary
                                }

The intracluster gas in Abell 2199, Abell 3571 and Abell 496 show
significant central metal abundance enhancements.
In Abell 2199 the abundance declines from $0.49 \pm 0.04$ solar at the
center to $0.34 \pm 0.03$ solar in the outer parts ($>3'$).
The central abundance enhancement in Abell 3571 is weaker, but still
significant: the abundance is $0.37\pm 0.06$ solar at the center and
declines to $0.28\pm 0.03$ solar beyond $3'$.
The abundance in Abell 496 declines from $0.53 \pm 0.04$ solar near the
center to $0.36 \pm 0.03$ solar beyond $>3'$ (Dupke \& White 1999).

The central regions of Abell 2199, Abell 3571 and Abell 496 are significantly
enriched by SN Ia ejecta.
A variety of abundance ratios were used to show that the SN Ia iron
mass fractions near their centers are 74$\pm 7\%$, 58$\pm 17\%$
73$\pm 5\%$, respectively.

The Ni/Fe ratios observed in all three clusters are $inconsistent$ with the predicted
values from the Nomoto et al.\ (1997b) delayed detonation models for SN Ia, at
greater than the 90\% confidence level.
However, the predicted Ni/Fe ratio of the updated W7 deflagration model
(Nomoto et al.\ 1997b) is $consistent$ with the observed values.
These results illustrate the power of X-ray cluster spectroscopy
to constrain theoretical models of supernovae.
Forthcoming observations with {\sl Chandra}, {\sl XMM} and {\sl Astro-E}
should provide higher quality constraints using these and additional
abundance ratios.

                        \acknowledgments

This work was partially supported by the NSF and the State of Alabama through
EPSCoR grant EHR-9108761.  REW also acknowledges partial support from NASA grant
NAG 5-2574. This research made use of the HEASARC {\sl ASCA} database and NED. We
would like to thank Dr. K. Arnaud for helpful discussions.

\clearpage

                                
                                \clearpage

                                \begin{figure}
                                \title{
Figure Captions
                                }

                                \caption{
Radial distributions of overall elemental abundances in Abell 496, Abell 2199
and Abell 3571, in solar units;
the abundance errors are 90\% confidence limits; the radial ``errors''
indicate the size of the spectral extraction regions.
                                }

                                \caption{
Comparison of the Ni/Fe abundance ratios in the inner projected regions ($0 - 2'$)
of Abell 496, Abell 2199 and Abell 3571. Theoretical predictions for
different SN Ia models, W7 (thick line) and
delayed detonation models WDD1, WDD2 and WDD3 (thin lines), are also shown.
                                }

                                \end{figure}
\clearpage

\begin{deluxetable}{lcccc}
\small
\tablewidth{0pt}
\tablecaption{Effective Exposure Times}
\tablehead{
\colhead{Spectrometer } &
\multicolumn{3}{c} {Exposure Time (ksec)} & \nl
\colhead{} &
\colhead{Abell 496}  &
\colhead{Abell 2199}  &
\colhead{Abell 3571}  &
}
\startdata
SIS 0 & 30 & 21 & 13 & \nl
SIS 1 & 23 & 16 & 13  & \nl
GIS 2 & 40 & 34 & 25 & \nl
GIS 3 & 40 & 34 & 25 & \nl
\enddata
\end{deluxetable}
 \clearpage
\begin{deluxetable}{lcccc}
\small
\tablewidth{0pt}
\tablecaption{Spectral Fits\tablenotemark{a,b}}
\tablehead{
\colhead{Cluster} &
\colhead{Region \tablenotemark{c}}  &
\colhead{$kT$}  &
\colhead{Abundance} &
\colhead{$\chi^2_\nu$}  \nl
\colhead{} &
\colhead{(arcmin)} &
\colhead{(keV)}  &
\colhead{(solar)} &
\colhead{}
}
\startdata
Abell 496     & $0 - 2$ & 3.24$^{+0.07}_{-0.06}$ & 0.53$^{+0.04}_{-0.04}$ & 1.03\nl
\hspace{4mm} $\prime \prime$  & $0 - 2^{*}$ & 3.37$^{+0.08}_{-0.09}$ & 0.54$^{+0.05}_{-0.05}$ & 1.03\nl
\hspace{4mm} $\prime \prime$  & $3 - 12$ & 4.28$^{+0.08}_{-0.08}$ & 0.36$^{+0.03}_{-0.03}$ & 1.07\nl
Abell 2199     & $0 - 2$ & 3.71$^{+0.09}_{-0.09}$ & 0.49$^{+0.04}_{-0.04}$ & 1.14\nl
\hspace{4mm} $\prime \prime$  & $0 - 2^{*}$ & 4.22$^{+0.14}_{-0.15}$ & 0.49$^{+0.04}_{-0.04}$ & 1.08\nl
\hspace{4mm} $\prime \prime$  & $3 - 12$ & 4.14$^{+0.06}_{-0.06}$ & 0.34$^{+0.03}_{-0.03}$ & 1.14\nl
Abell 3571     & $0 - 2$ & 6.45$^{+0.36}_{-0.33}$ & 0.37$^{+0.06}_{-0.05}$ & 1.09\nl
\hspace{4mm} $\prime \prime$  & $0 - 2^{*}$ & 8.4$^{+0.5}_{-0.9}$ & 0.42$^{+0.07}_{-0.06}$ & 1.04\nl
\hspace{4mm} $\prime \prime$  & $3 - 12$ & 6.89$^{+0.20}_{-0.19}$ & 0.28$^{+0.03}_{-0.03}$ & 1.24\nl

\enddata
\tablenotetext{a}{Errors are 90\% confidence limits}
\tablenotetext{b}{Simultaneous fittings of SIS 0 \& 1, GIS 2 \& 3}
\tablenotetext{c}{Distance from the X-ray center }
\tablenotetext{*}{same as above but with an extra  cooling flow component}
\end{deluxetable}
\clearpage
\begin{deluxetable}{lcccc}
\small
\tablewidth{0pt}
\tablecaption{Individual Elemental Abundances for the Inner Regions \tablenotemark{a}}
\tablehead{
\colhead{} &
\colhead{} &
\colhead{Cluster} &
\colhead{}\nl
\colhead{Element} &
\colhead{Abell 496 }  &
\colhead{Abell 2199 } &
\colhead{Abell 3571 }
}
\startdata
O & 0.48$^{+0.39}_{-0.37}$ & 0.65$^{+0.21}_{-0.20}$ & $\ge 0.9$ \nl
Ne & 0.87$^{+0.45}_{-0.40}$ & 0.97$^{+0.49}_{-0.46}$ & 1.87$^{+1.96}_{-1.56}$  \nl
Si & 0.83$^{+0.18}_{-0.17}$ & 0.59$^{+0.21}_{-0.20}$ & 1.29$^{+0.85}_{-0.71}$ \nl
S & 0.58$^{+0.19}_{-0.20}$ & 0.23$^{+0.25}_{-0.22}$ & 0.40$^{+0.90}_{-0.40}$ \nl
Fe & 0.53$^{+0.05}_{-0.05}$ & 0.49$^{+0.05}_{-0.05}$ & 0.46$^{+0.09}_{-0.08}$ \nl
Ni & 2.57$^{+0.67}_{-0.80}$ & 1.79$^{+0.89}_{-0.85}$ & 2.66$^{+1.82}_{-1.64}$ \nl
\enddata
\tablenotetext{a}{Errors are 90\% confidence limits}
\end{deluxetable}
  \clearpage

\begin{deluxetable}{lcccccc}
\small
\tablewidth{0pt}
\tablecaption{Elemental Abundance Ratios\tablenotemark{a}}
\tablehead{
\colhead{Element} &
\multicolumn{3}{c} {Central Region} &
\multicolumn{3}{c} {SN Ia Theory\tablenotemark{b}} \nl
\colhead{Ratio} &
\colhead{Abell 496 }  &
\colhead{Abell 2199 } &
\colhead{Abell 3571} &
\colhead{W7} &
\colhead{WDD2} &
\colhead{SN II}
}
\startdata
O/Fe & 0.91$^{+0.44}_{-0.44}$ & 1.33$^{+0.85}_{-0.79}$ &  --- & 0.037 & 0.019 & 3.82   \nl
Si/Fe & 1.57$^{+0.22}_{-0.22}$ & 1.20$^{+0.26}_{-0.26}$ & 2.8$^{+1.17}_{-1.09}$ &  0.538 & 1.013 & 3.53   \nl
Ni/Fe & 4.85$^{+1.02}_{-0.96}$ & 3.65$^{+1.27}_{-1.09}$ & 5.8$^{+2.61}_{-2.43}$ &  4.758 & 1.4 & 1.65  \nl
O/Si & 0.58$^{+0.72}_{-0.29}$ & 1.10$^{+0.73}_{-0.68}$ &  --- & 0.068 & 0.019 & 1.1   \nl
Si/Ni & 0.32$^{+0.08}_{-0.08}$ & 0.33$^{+0.13}_{-0.12}$ & 0.48$^{+0.28}_{-0.26}$ &  0.113 & 0.725 & 2.14  \nl
\enddata
\tablenotetext{a}{Errors are  propagated 1$\sigma$ errors}
\tablenotetext{b}{SN Ia: Nomoto et al (1997a); SN II: Nomoto et al (1997b)}
\end{deluxetable}
\clearpage
\begin{deluxetable}{lcccc}
\small
\tablewidth{0pt}
\tablecaption{Fe Mass Fraction from Different Ratios\tablenotemark{a}}
\tablehead{
\colhead{Element} &
\multicolumn{3}{c}{SN Ia Iron Mass Fraction} \nl
\colhead{Ratio} &
\colhead{Abell 496} &
\colhead{Abell 2199} &
\colhead{Abell 3571}
}
\startdata
O/Fe & 0.77$^{+0.13}_{-0.13}$ & 0.66$^{+0.21}_{-0.23}$ &  ---  \nl
Si/Fe & 0.65$^{+0.07}_{-0.08}$ & 0.77$^{+0.09}_{-0.08}$ & 0.25$^{+0.36}_{-0.25}$  \nl
Ni/Fe & 1.00$^{+0.00}_{-0.27}$ & 0.66$^{+0.34}_{-0.35}$ & 1.00$^{+0.00}_{-0.45}$  \nl
O/Si & 0.86$^{+0.10}_{-0.86}$ & $\le 0.92$ &  ---   \nl
Si/Ni & 0.74$^{+0.07}_{-0.09}$ & 0.74$^{+0.13}_{-0.11}$ & 0.61$^{+0.25}_{-0.18}$  \nl
$Average$ &0.73$\pm 0.05$ & 0.74$\pm 0.07$ & 0.58$\pm 0.17$ \nl
\enddata
\tablenotetext{a}{Errors are  propagated 1$\sigma$ errors}
\tablenotetext{b}{SN Ia: Nomoto et al (1997a); SN II: Nomoto et al (1997b)}
\end{deluxetable}
\clearpage
\begin{deluxetable}{lccccc}
\small
\tablewidth{0pt}
\tablecaption{Iron Mass Fraction from Different Delayed Detonation Models\tablenotemark{a}}
\tablehead{
\colhead{Cluster} &
\colhead{Element} &
\multicolumn{3}{c}{SN Ia Iron Mass Fraction\tablenotemark{b}} \nl
\colhead{} &
\colhead{Ratio} &
\colhead{WDD1} &
\colhead{WDD2} &
\colhead{WDD3}
}
\startdata
A496 & O/Fe & 0.82$^{+0.11}_{-0.10}$ & 0.77$^{+0.12}_{-0.11}$ & 0.74$^{+0.12}_{-0.12}$   \nl
\hspace{4mm} $\prime \prime$   & Si/Fe & $\ge 0.96$ & 0.79$^{+0.08}_{-0.09}$ & 0.65$^{+0.08}_{-0.08}$  \nl
\hspace{4mm} $\prime \prime$  & Ni/Fe &  ---  &  ---  &  ---  \nl
\hspace{4mm} $\prime \prime$   & O/Si & 0.72$^{+0.17}_{-0.72}$ & 0.77$^{+0.15}_{-0.77}$ & 0.81$^{+0.12}_{-0.81}$   \nl
\hspace{4mm} $\prime \prime$   & Si/Ni &  ---  &  ---  &  ---  \nl
\hspace{4mm} $\prime \prime$   & $WA\tablenotemark{c}$ & 0.82$^{+0.11}_{-0.10}$ & 0.78$\pm 0.07$ & 0.68$\pm 0.07$ \nl
\hline
A2199 & O/Fe & 0.72$^{+0.18}_{-0.21}$ & 0.66$^{+0.22}_{-0.21}$ & 0.62$^{+0.22}_{-0.22}$   \nl
\hspace{4mm} $\prime \prime$   & Si/Fe &  ---  & 0.93$^{+0.10}_{-0.10}$ & 0.78$^{+0.10}_{-0.09}$  \nl
\hspace{4mm} $\prime \prime$   & Ni/Fe &  ---  &  ---  &  ---   \nl
\hspace{4mm} $\prime \prime$   & O/Si & $\le 0.83$ & $\le 0.86$ & $\le 0.89$   \nl
\hspace{4mm} $\prime \prime$   & Si/Ni &  ---  &  ---  & $\ge 0.96$  \nl
\hspace{4mm} $\prime \prime$   & $WA$ &0.72$^{+0.18}_{-0.21}$ & 0.88$\pm 0.09$ & 0.75$\pm 0.09$ \nl
\hline
A3571 & O/Fe &  --- &  --- &  ---  \nl
\hspace{4mm} $\prime \prime$   & Si/Fe & 0.47$^{+0.52}_{-0.47}$ & 0.30$^{+0.43}_{-0.30}$ & 0.23$^{+0.37}_{-0.23}$  \nl
\hspace{4mm} $\prime \prime$   & Ni/Fe &  ---  &  ---  &  ---  \nl
\hspace{4mm} $\prime \prime$   & O/Si &  --- &  --- &  ---   \nl
\hspace{4mm} $\prime \prime$   & Si/Ni &  ---  & $\ge 0.98 $ & 0.95$^{+0.17}_{-0.17}$  \nl
\hspace{4mm} $\prime \prime$   & $WA$ & 0.47$^{+0.52}_{-0.47}$ & 0.30$^{+0.43}_{-0.30}$ & 0.82$\pm 0.15$ \nl
\enddata
\tablenotetext{a}{Errors are  propagated 1$\sigma$ errors}
\tablenotetext{b}{SN Ia: Nomoto et al (1997a); SN II: Nomoto et al (1997b)}
\tablenotetext{c}{Weighted Average}
\end{deluxetable}

                                
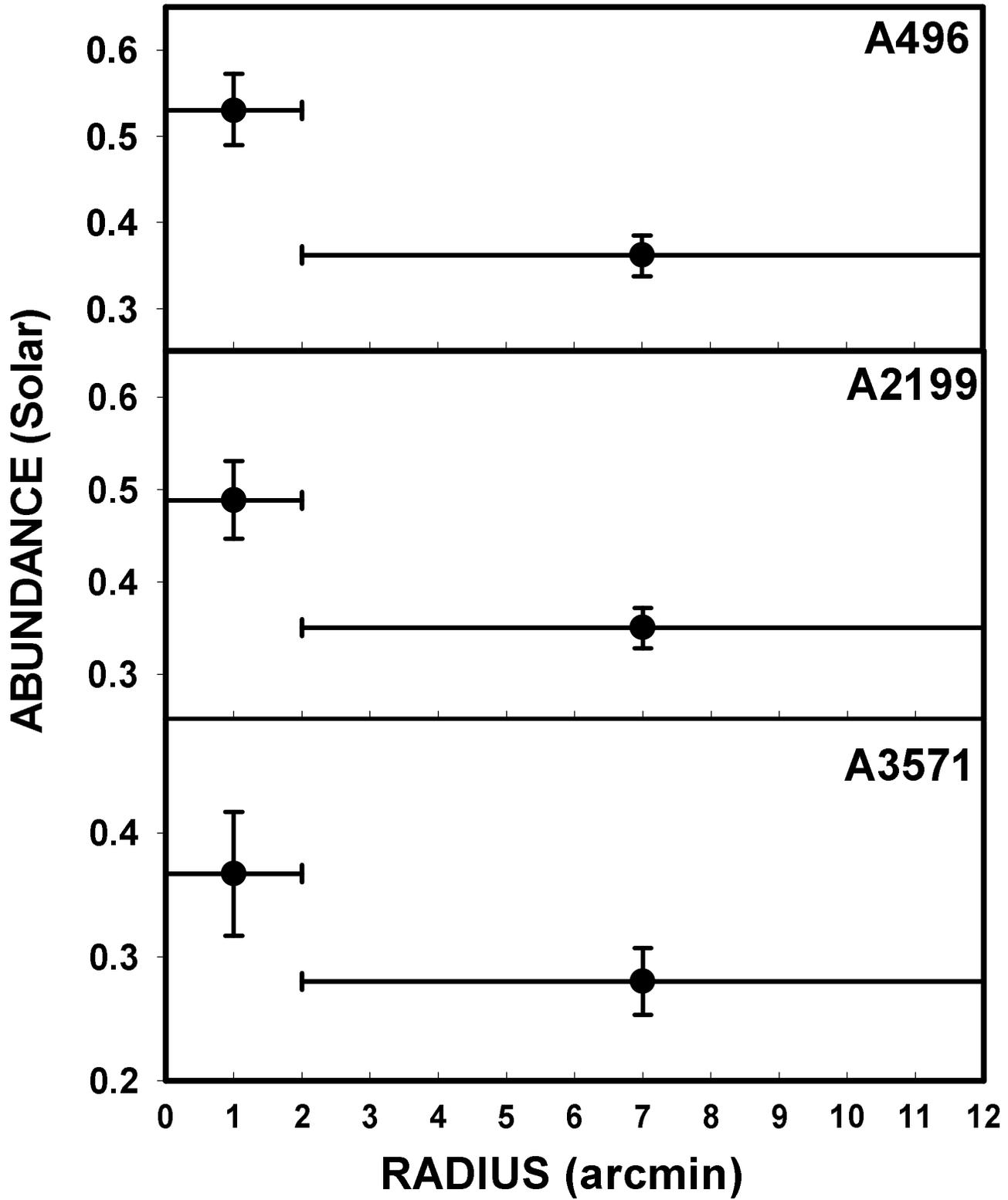
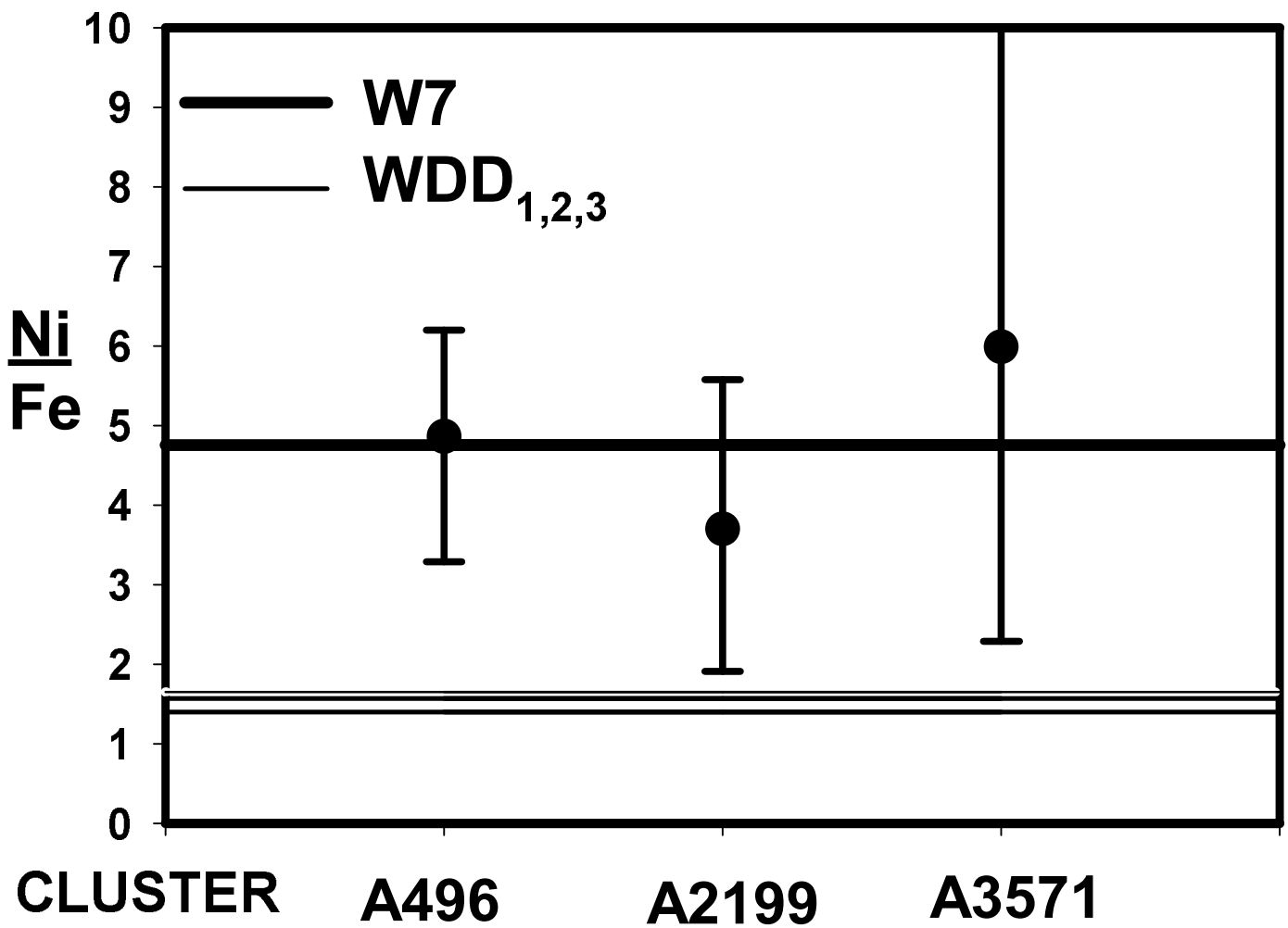
\end{document}